\documentclass[journal]{IEEEtran}
\usepackage{amsmath,amsfonts}
\usepackage{array}
\usepackage[caption=false,font=normalsize,labelfont=sf,textfont=sf]{subfig}
\usepackage{textcomp}
\usepackage{stfloats}
\usepackage{url}
\usepackage{verbatim}
\usepackage{graphicx}
\usepackage{cite}

\usepackage{diagbox}
\usepackage{physics}

\usepackage{algpseudocode}
\usepackage{algorithm}
\usepackage{hyperref}

\newcommand*\BitwiseAND{\mathbin{\&}}
\newcommand*\LogicalAND{\mathbin{\&\&}}
\newcommand*\LogicalOR{\mathbin{||}}
\newcommand*\BitwiseOR{\mathbin{|}}
\newcommand*\ShiftLeft{\ll}
\newcommand*\ShiftRight{\gg}
\newcommand*\LogicalNOT{!}

\allowdisplaybreaks
\begin{document}

\title{New Difference Triangle Sets by an\\ {FPGA}-Based Search Technique}

\author{Mohannad~Shehadeh, William~Kingsford, and
	Frank~R.~Kschischang
	\thanks{Mohannad~Shehadeh and Frank~R.~Kschischang are with the Edward S. Rogers Sr. Department of Electrical \& Computer Engineering, University of Toronto, Toronto, ON M5S 3G4, Canada (emails: \{mshehadeh, frank\}@ece.utoronto.ca).}
	\thanks{William~Kingsford is an unaffiliated scholar from Palmerston North, New Zealand (email: will.kingsford@gmail.com).}
}

\maketitle

\begin{abstract}
	We provide some difference triangle sets with scopes
	that improve upon the best known values.
	These are found with purpose-built digital circuits
	realized with field-programmable gate
	arrays (FPGAs) rather than software algorithms
	running on general-purpose processors.
\end{abstract}


\section{Introduction}

\IEEEPARstart{D}{ifference} triangle sets (DTSs) have long been studied in information theory \cite{Klove,Klove-Bounds,Colbourn-IT,Zhi-DDS,Chen-Further,DTS-Affine,Koubi-Hill-Climbing,MILP-DTS,Some-DTS} due to their
applications to various problems of information processing
and transmission. These include applications to the design of error-correcting codes \cite{CSOC,NB-DTS,CC-DTS,hosc}, code-division multiple-access (CDMA) codes \cite{OOC,DTS-SOOC,SOOC}, 
and various codes for interference 
and ambiguity management in measurement, imaging, and communication systems \cite{Bermond,Golomb-Ruler,Babcock}.

An $(n,k)$-DTS is an $n \times (k+1)$ array of 
integers, $A$ with entries 
$a_{ij}$ for $i \in \{0,1,\dots,n-1\}$
and $j \in \{0,1,\dots,k\}$, such that 
all differences $a_{ij}-a_{ij'}$ where
$0\leq i \leq n-1$ and $0 \leq j \neq j' \leq k$
are \emph{distinct} and \emph{nonzero}. 
The \emph{scope} of an $(n,k)$-DTS $A$, denoted
$m(A)$, is its largest such difference
\begin{equation*}
	m(A) = \max \{a_{ij}-a_{ij'}\mid 0\leq i \leq n-1, 0 \leq j \neq j' \leq k\}\text{.}
\end{equation*}
An $(n,k)$-DTS $A$ is said to be \emph{normalized} if 
\begin{equation*}
	0 = a_{i0} < a_{i1} < \cdots < a_{ik}
\end{equation*}
for all $i \in \{0,1,\dots,n-1\}$ in which case the scope
of $A$ is simply its largest element
\begin{equation*}
	m(A) = \max \{a_{ij}\mid 0\leq i \leq n-1, 0 \leq j \leq k\}\text{.}
\end{equation*}
We will assume henceforth that all DTSs are normalized. 
We also describe DTSs with geometric language
where the rows of a DTS are called \emph{rulers}, 
the entries are called \emph{marks}, and the positive
differences between marks are called \emph{distances}.

In the applications of DTSs, a central problem
is that of exhibiting DTSs with minimum scope, referred to
as optimal DTSs. This is a difficult
problem which generalizes the widely studied 
optimal Golomb ruler problem \cite{distributed-OGR,Golomb-Ruler,Apostol-Golomb} where
a \emph{Golomb ruler} is a $(1,k)$-DTS. The key quantity
of interest is then the minimum possible scope
for an $(n,k)$-DTS denoted $m(n,k)$ where
\begin{equation*}
	m(n,k) = \min\{m(A) \mid \text{$A$ is an $(n,k)$-DTS}\}\text{.}
\end{equation*}

We have a trivial lower bound on $m(n,k)$ given
by
\begin{equation*}
	m(n,k) \geq n \cdot \frac{k(k+1)}{2}
\end{equation*}
and holding with equality if and only if the
set of distances $a_{ij} - a_{ij'}$
where $0 \leq i \leq n-1$ and $0 \leq j' < j \leq k$
is precisely $\{1,2,\dots,nk(k+1)/2\}$. A DTS for
which the trivial lower bound
holds with equality is said to be \emph{perfect} and is necessarily optimal.

Except for when $k = 1$
and $k = 2$, it is an open problem to 
exhibit minimum scope $(n,k)$-DTSs for arbitrary $n$.
When $k=3$ or $k=4$, certain recursive constructions
establish the existence of perfect DTSs 
for infinitely many---but far from all---values of $n$.
Apart from these constructions not covering 
all values of $n$, it is also known that perfect DTSs
can only exist for $k \leq 4$ \cite{Colbourn-IT}. 
Beyond these cases,
algebraic constructions can be used
to obtain good upper bounds on
$m(n,k)$, but heuristic algorithms
appear to be the best methods for finding
the smallest scope DTSs and thus 
the best upper bounds on $m(n,k)$ for specific
values of $n$ and $k$. In addition to their
direct use in practical applications, 
good or optimal DTSs obtained
by heuristic methods can also be theoretically
useful, for example, as base cases in recursive constructions.

Our contributions in this paper are exhibiting 
better DTSs than the best
known DTSs for various parameters 
and the introduction of the new, 
highly unconventional computational search technique
used to find them. This entails using
custom digital circuits that are realized with 
programmable hardware technology to subvert the
computing models that are imposed by general-purpose processors.
While unconventional, we will find that this is a very natural 
approach to computationally challenging combinatorial design problems.
The remainder of this paper is organized
as follows: In Section \ref{background-section},
we briefly survey existing key results and methods.
Section \ref{results-section} provides our
improved upper bounds on $m(n,k)$ with the corresponding
DTSs exhibited in Appendix \ref{DTS-appendix}. Lastly, our search technique
is described in Section \ref{search-technique-section}
and we conclude in Section \ref{conclusion-section} with
some suggestions for future work. Details of the
devices used and performance estimates 
are provided in Appendix \ref{engineering-appendix}.

\section{Background}\label{background-section}

\subsection{Key Results for Small $k$}

We review some key results for small $k$.
When $k = 1$, the trivial lower bound is
\begin{equation}\label{k=1}
	m(n,1) \geq n
\end{equation}
and is trivially achieved by taking $a_{i0} = 0$ and $a_{i1} = i + 1$
for $0 \leq i \leq n-1$. When $k = 2$, the trivial lower bound is
slightly strengthened to
\begin{equation}\label{k=2}
	m(n,2) \geq 
	\begin{cases}
		3n & \text{if } n \equiv 0 \text{ or } 1 \pmod{4} \\
		3n + 1 & \text{if } n \equiv 2 \text{ or } 3\pmod{4}
	\end{cases}
\end{equation}
by considering the parity of the sum
of all distances.
This is achieved for arbitrary $n$ by the explicit
constructions of Skolem \cite{Skolem} and 
O'Keefe \cite{OKeefe}, which respectively cover the two cases, 
via a well-known equivalence of $(n,2)$-DTSs 
to the objects considered therein. The 
inequalities \eqref{k=1} and \eqref{k=2}
are therefore exact equalities.

When $k = 3$, the trivial lower bound is
\begin{equation}
	m(n,3) \geq 6n
\end{equation}
and it is a longstanding conjecture 
of Bermond \cite{Bermond} that this is exact
for all $n$ excluding $n\in \{2,3\}$. 
Certain recursive constructions such 
as that of Wild \cite{Wild} due to the results
of Kotzig and Turgeon \cite{Kotzig-Turgeon}
can be used to show that there
are infinitely many values of $n$ for which an
$(n,3)$-DTS of scope $6n$ exists.
Bermond's conjecture has been verified
for all $n \leq 1000$ by Ge et al.\ in \cite{deg-4-1000}
by using a wide variety of recursive constructions
to cover gaps as needed.

When $k = 4$, the trivial bound again slightly strengthens
to
\begin{equation}\label{k=4}
	m(n,4) \geq 
	\begin{cases}
		10n & \text{if } n \equiv 0 \pmod{2} \\
		10n + 1 & \text{if } n \equiv 1 \pmod{2}
	\end{cases}
\end{equation}
by again considering the parity of the sum of all distances.
This is conjectured to be exact for all $n$ excluding $n\in\{2,3,4\}$.
The recursive construction of Wild \cite{Wild} implies 
that there are infinitely many values of $n$ for which 
an $(n,4)$-DTS of scope $10n$ exists. However, not many
examples exist for small values of $n$. We extend
the verification of this conjecture to $n = 15$ in this
paper.

When $k > 4$, much less is known. Improved lower bounds
on $m(n,k)$
are provided by Kl{\o}ve in \cite{Klove}. Better 
lower bounds are provided by Lorentzen and Nilsen via linear 
programming in \cite{MILP-DTS} 
and these are further improved by Shearer in \cite{Shearer-LP}. 
Lastly, several small cases are settled
by an exhaustive backtracking search program of Shearer
\cite{Shearer-IBM-Programs-Paper,Shearer-IBM-Programs}.

\subsection{Existing Search Techniques}

Accounting for obvious invariants, 
a naive exhaustive 
search for $(n,k)$-DTSs requires 
choosing $n$ sets of rows from a set of 
at least $\binom{nk(k+1)/2}{k}$
possible rows due to the trivial lower bound 
on scope. This leads to
a search space size of at least
\begin{equation*}
	\binom{\binom{nk(k+1)/2}{k}}{n}
\end{equation*}
which is astronomical.

Naturally, many lines of work \cite{Colbourn-IT,TABU,Koubi-Hill-Climbing,Koubi-Followup}
consider heuristic algorithms for combinatorial optimization
including greedy algorithms, local search, and 
tabu search. Other lines of work \cite{CSOC,Klove,Chen-Further,DTS-Affine,Shearer-IBM-Report} consider
algebraic constructions that are typically based on 
modular Golomb rulers \cite[Part~VI:~Ch.~19]{Colbourn}.
These methods are also heavily computer- and heuristic-aided,
amounting to a reduction of the search space to good 
structured classes. Nonetheless, they are quite powerful as demonstrated by Shearer who
extends and combines such methods with computer 
search heuristics to find many of the best upper bounds
on $m(n,k)$ in \cite{Shearer-IBM-Report}.

\subsection{Equivalence to Strict Optical Orthogonal Codes}

Optical orthogonal codes (OOCs) \cite{OOC} 
are CDMA codes that
are used with unipolar $\{0,1\}$-valued 
signaling---occurring in non-coherent 
optical communication---as opposed
to bipolar $\{1,-1\}$-valued signaling. 
As a result, they are defined in terms of the
correlation properties of $\{0,1\}$-valued sequences.
An 
$(N,w)$-OOC is a set $\mathcal{C}\subseteq \{0,1\}^N$ 
of length $N$ sequences of weight $w$ 
with the following autocorrelation
and cross-correlation properties:
\begin{itemize}
	\item For all codewords $x\in \mathcal{C}$ where
	$x = (x_0,x_1,\dots,x_{N-1})$,
	\begin{equation*}
		\sum_{t=0}^{N-1}x_t x_{t\oplus\tau} \leq 1
	\end{equation*}
	for all nonzero integers $\tau$ where $\oplus$ denotes modulo-$N$ addition.
	\item For all codeword pairs $x,y\in \mathcal{C}$ where
	$x \neq y$,
	\begin{equation*}
		\sum_{t=0}^{N-1}x_t y_{t\oplus\tau} \leq 1
	\end{equation*}
	for all integers $\tau$.
\end{itemize}

By interpreting the rows of an $(n,k)$-DTS $A$ as defining
the locations of $w = k + 1$ nonzero entries of $n$ codewords, we get a constant weight code $\mathcal{C}\subseteq \{0,1\}^N$
where $\abs{\mathcal{C}} = n$ and $N \geq m(A)+1$. If we take
$N = 2m(A)+1$, it can be shown that the resulting code
is a $(2m(A)+1,k+1)$-OOC.
Taking $N=2m(A)+1$ as opposed to $N=m(A)+1$ is required to preserve
modulo-$N$ distinctness of differences which will
arise as a requirement when deducing this fact.

While not every OOC can be obtained from a DTS in this way,
Chu and Golomb establish in \cite{DTS-SOOC} that a special
family of OOCs introduced by Zhang in \cite{SOOC} termed \emph{strict} OOCs (S-OOCs), are precisely equivalent to those obtained from 
DTSs as just described. Compared to conventional OOCs \cite{OOC}, S-OOCs are robust
to fully asynchronous communication as shown in \cite{SOOC}. 
Importantly, the minimum scope
DTS problem considered in this paper is equivalent to the minimum
length S-OOC problem of \cite{SOOC} and our improved (smaller scope) 
DTSs translate to improved (smaller length) S-OOCs.

\section{Results}\label{results-section}

\begin{table*}[!t]
	\caption{Improved Upper Bounds for $m(n,k)$\label{main-table}} 
	\centering
	\resizebox{\textwidth}{!}{
	\begin{tabular}{c|ccccccccccccccc}
		\diagbox{$k$}{$n$} & $4$ & $5$ & $6$ & $7$ & $8$ & $9$ & $10$ & $11$ & $12$ & $13$ & $14$ & $15$ \\ \hline
		$4$ & & & & & & & & &
		$\underset{\displaystyle (122)}{120}$ & 
		$\underset{\displaystyle (133)}{131}$ & 
		$\underset{\displaystyle (143)}{140}$ &
		$\underset{\displaystyle (154)}{151}$ \\ 
		$5$ & & & &
		$\underset{\displaystyle (113)}{112}$ &
		$\underset{\displaystyle (129)}{128}$ &
		$\underset{\displaystyle (146)}{145}$ &	
		$\underset{\displaystyle (164)}{163}$ & 
		$\underset{\displaystyle (181)}{180}$ & 
		$\underset{\displaystyle (197)}{196}$ & 
		$\underset{\displaystyle (214)}{213}$ &
		$\underset{\displaystyle (231)}{230}$ &
		$\underset{\displaystyle (250)}{249}$    \\ 
		$6$ & & 
		$\underset{\displaystyle (119)}{118}$ & 
		$\underset{\displaystyle (145)}{144}$ &
		$\underset{\displaystyle (171)}{170}$ &
		$\underset{\displaystyle (197)}{196}$ & 
		$\underset{\displaystyle (222)}{221}$ &
		$\underset{\displaystyle (248)}{247}$ &
		$\underset{\displaystyle (274)}{273}$ &
		$\underset{\displaystyle (299)}{298}$ &
		$\underset{\displaystyle (324)}{322}$ &
		$\underset{\displaystyle (352)}{351}$ & 
		$\underset{\displaystyle (374)}{373}$ \\ 
		$7$ & 
		$\underset{\displaystyle (135)}{134}$ & 
		$\underset{\displaystyle (171)}{170}$ & 
		$\underset{\displaystyle (210)}{208}$ & 
		$\underset{\displaystyle (251)}{249}$ & 
		$\underset{\displaystyle (282)}{281}$ & 
		$\underset{\displaystyle (318)}{317}$ & 
		$\underset{\displaystyle (352)}{351}$ & 
		$\underset{\displaystyle (393)}{391}$ & 
		$\underset{\displaystyle (431)}{430}$ & 
		$\underset{\displaystyle (464)}{461}$ & 
		$\underset{\displaystyle (499)}{498}$ & 
		$\underset{\displaystyle (524)}{523}$ & 		
	\end{tabular}
	}
\end{table*}

Table \ref{main-table} provides our improved upper bounds on $m(n,k)$
with the corresponding $(n,k)$-DTSs provided explicitly in Appendix \ref{DTS-appendix}. 
The previous
best values of $m(n,k)$ are shown in parentheses underneath our improved values.
These previous best upper bounds are taken from the 
compiled results of \cite[Part~VI:~Ch.~19]{Colbourn} together with the
results of
\cite{Shearer-IBM-Report, Koubi-Hill-Climbing, Koubi-Followup, TABU}
which include some more recent improvements. Note that the 
$(14,4)$-DTS and the $(15,4)$-DTS provided are optimal due
to the lower bound \eqref{k=4}. Moreover,
a software version of our proposed search
technique was used in \cite{hosc} to find an
optimal $(12,4)$-DTS and $(13,4)$-DTS 
which are provided therein.

\section{Search Technique}\label{search-technique-section}

We begin with a brief overview of compute devices
and programmable hardware in Section \ref{FPGA-background-section}. 
We follow this with a description of a
software version of our algorithm in Section \ref{software-section}
and an overview of the hardware version in Section \ref{hardware-section}.
Both the software and hardware implementations are made available online
in full \cite{dts-search-hdl}.

\subsection{On Compute Devices and Programmable Hardware}\label{FPGA-background-section}

Our search technique employs \emph{field-programmable gate arrays (FPGAs)}
as opposed to \emph{processors} such as central processing
units (CPUs) and graphics processing units (GPUs).
Processors execute predefined instructions and are programmed
with software which translates to sequences of such instructions. 
On the other hand, 
FPGAs are programmed with digital circuit descriptions that are
specified with a \emph{hardware-description language (HDL)}. 
These descriptions are
translated into configuration files that effectively rewire 
the FPGA to realize the desired logic. Digital circuits
can be specified at the level of logic gates, flip-flops, 
and the connections between them but are typically specified
with a slightly higher abstraction termed \emph{register-transfer level (RTL)}. 
The reader is referred to \cite{Brown-Vranesic} for further
background on modern digital design.

The ability of an FPGA to 
realize arbitrary digital circuits 
does not at all mean that an FPGA can outperform a CPU or a GPU
on any given task.
This is because the reconfigurability of FPGAs comes at a significant
cost to both area and speed relative to hardware that is not
reconfigurable \cite{Kuon-Rose}, of which processors are examples.
Modern processors already excel at
arithmetic-heavy workloads with CPUs being tailored for complex, branching operations and GPUs being
tailored for simpler, massively-parallel operations.
Moreover, modern CPUs also support a significant amount
of parallelism both through single-instruction multiple-data (SIMD) 
capabilities and multithreading. Getting an 
FPGA speedup relative to a processor thus
requires that a smaller amount of slower
but arbitrarily configurable logic can 
outperform the significantly larger and faster 
but predefined logic provided by a processor. 
This occurs when there is a big mismatch between
the type of processing that is required and the 
compute model, compute units, and data flows 
that are afforded by that processor.

Our proposed technique benefits from
FPGAs because it involves a significant amount of custom
logic but very little arithmetic and very little branching.
In addition, we rely on unusual 
wide, shallow memories and wide, short datapaths with
feedback paths which is very unlike conventional 
processors. 
This results in a best-case scenario where our
DTS search
units are individually faster than a single CPU thread executing
a software version of the search algorithm
\emph{and} they are small enough for a large number of them to 
be fit onto a single FPGA. We speculate that
other computationally challenging combinatorial design problems
might admit similar hardware approaches since they tend to have
small discrete spatial representations that map naturally
onto circuits.

\subsection{Software Implementation}\label{software-section}

Our hardware design implements
a hardware-friendly variation of the 
DTS search algorithm of Koubi et al.\ \cite{Koubi-Hill-Climbing}.
Our implementation 
is largely based on the variation
of this algorithm that we provide in a previous
work \cite{hosc} where it was implemented
as optimized software making use
of a standard programming trick 
in which a subset of $\{0,1,\dots,M\}$ is represented
by a single $(M+1)$-bit integer. We will
describe this approach, with much of what follows
being replicated from \cite{hosc}.

The algorithm of Koubi et al.\ \cite{Koubi-Hill-Climbing} involves 
populating a partially constructed DTS mark by mark. 
As noted in \cite{hosc}, the innermost
loop step of any such algorithm is checking
whether a new mark fits into our partially constructed DTS
without inducing repeated distances. This requires a large
number of distance calculations and comparisons which
we can avoid by using \emph{bitmask} representations
and employing Algorithm \ref{fast-conditional-mark-insertion}
which we replicate from \cite{hosc}.
We describe this approach with $C$-language-like
\cite{C-language} notation. Note that the Verilog HDL \cite{Verilog} 
uses a similar syntax making it easy to relate this
to our provided software and hardware implementations \cite{dts-search-hdl}. We review key elements of 
this notation for the reader's convenience:
\begin{itemize}
	\item An $(M+1)$-bit integer variable $\mathsf{x}$ is
	interpreted as an element of the ring of integers
	modulo $M+1$ and its $i$th bit is the value
	of $\mathsf{b}_i \in \{0,1\}$ where $\mathsf{x} = \sum_{i=0}^{M} \mathsf{b}_i2^i$.
	\item The binary operators $\BitwiseAND$, $\BitwiseOR$, $\ShiftLeft$, and $\ShiftRight$ denote bitwise AND, bitwise OR, left shift, and right shift respectively.
	\item Nonzero integers are interpreted as logically true and zero as logically false when acted upon by the non-bitwise
	binary operators $\LogicalAND$ and $\LogicalOR$ and the unary operator $\LogicalNOT$ respectively denoting AND, OR, and NOT.
\end{itemize}

In Algorithm \ref{fast-conditional-mark-insertion}, 
variables can be interpreted as sets where a variable
represents the subset of $\{0,1,\dots,M\}$ containing 
all $i$ such that the $i$th bit of the variable is $1$.
A \emph{partially constructed DTS} is a set of possibly
incomplete rulers such that all distances between 
marks within a ruler are distinct from each other
and distinct from distances produced by other rulers.
Algorithm \ref{fast-conditional-mark-insertion} simply
inserts a given $\mathsf{mark}$ into a particular ruler $\mathsf{natRuler}$ if it does not induce repeated distances
and leaves the ruler unchanged otherwise. To do this
efficiently, we maintain:
\begin{itemize}
	\item the set of distances used thus
	far, by both the current ruler $\mathsf{natRuler}$ and all previously constructed rulers, denoted $\mathsf{usedDistances}$;
	\item the largest mark of $\mathsf{natRuler}$ denoted $\mathsf{largestMark}$; and
	\item a redundant mirror image of $\mathsf{natRuler}$ denoted $\mathsf{revRuler}$
	where $\mathsf{revRuler}$ contains $i$ if and only if $\mathsf{natRuler}$ contains 
	$\mathsf{largestMark}-i$.
\end{itemize}
The reader is referred to \cite{hosc} for further details if
the validity of Algorithm \ref{fast-conditional-mark-insertion}
is not clear.

\begin{figure*}[!t]
	\centering
	\includegraphics[width=\textwidth]{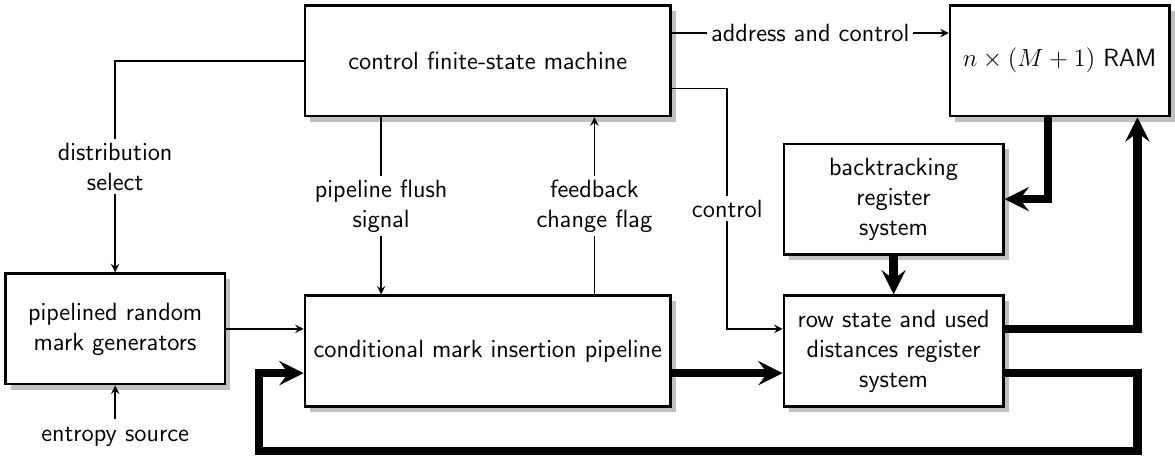}
	\caption{Simplified depiction of DTS search worker architecture; bold arrows indicate width $M+1$ internal datapaths.}\label{hardware-figure}
\end{figure*}

\begin{algorithm}[t]
	\caption{Fast conditional mark insertion \cite{hosc}.}\label{fast-conditional-mark-insertion}
	\begin{algorithmic}[1]
		\If{$\mathsf{mark} > \mathsf{largestMark}$} 
		\State $\mathsf{leftDistances} \gets \mathsf{revRuler} \ShiftLeft (\mathsf{mark}-\mathsf{largestMark})$
		\State $\mathsf{rightDistances} \gets 0$
		\Else
		\State $\mathsf{leftDistances} \gets \mathsf{revRuler} \ShiftRight (\mathsf{largestMark}-\mathsf{mark})$
		\State $\mathsf{rightDistances} \gets \mathsf{natRuler} \ShiftRight \mathsf{mark}$
		\EndIf
		\State $\mathsf{bisection} \gets \mathsf{leftDistances} \BitwiseAND \mathsf{rightDistances}$
		\State $\mathsf{distances} \gets \mathsf{leftDistances} \BitwiseOR \mathsf{rightDistances}$
		\State $\mathsf{intersection} \gets \mathsf{usedDistances} \BitwiseAND \mathsf{distances}$
		\If{$\LogicalNOT(\mathsf{bisection} \LogicalOR \mathsf{intersection})$}
		\State $\mathsf{natRuler} \gets \mathsf{natRuler} \BitwiseOR (1\ShiftLeft \mathsf{mark})$ 
		\If{$\mathsf{mark} > \mathsf{largestMark}$}
		\State $\mathsf{revRuler}\gets \mathsf{leftDistances} \BitwiseOR 1$
		\State $\mathsf{largestMark}\gets\mathsf{mark}$
		\Else
		\State $\mathsf{revRuler} \gets \mathsf{revRuler} \BitwiseOR (1\ShiftLeft (\mathsf{largestMark}-\mathsf{mark}))$ 
		\EndIf
		\State $\mathsf{usedDistances}\gets \mathsf{usedDistances} \BitwiseOR \mathsf{distances}$
		\EndIf
	\end{algorithmic}
\end{algorithm}

We then perform a DTS search with Algorithm \ref{modified-Koubi} which
is a modified version of the algorithm of Koubi et al.\ \cite{Koubi-Hill-Climbing}. This can be interpreted as
a greedy algorithm with simple backtracking steps: 
When a row cannot be completed after $\mathsf{thresh_2}$ attempts, 
we attempt to replace a previously completed row instead;
when the DTS cannot be completed after $\mathsf{thresh_1}$ attempts, it is abandoned.
Note that some additional record-keeping is required to do this
efficiently beyond what is presented in Algorithms \ref{fast-conditional-mark-insertion} and \ref{modified-Koubi},
which we omit for clarity.

A key component of the algorithm
of Koubi et al.\ \cite{Koubi-Hill-Climbing}
is that marks are not sampled uniformly:
The marks of a ruler in a DTS 
are modeled as arising from
Gaussian distributions where
the $j$th mark of any ruler 
in the DTS is sampled independently 
from a Gaussian distribution 
with mean $\mu_j$ and variance $\sigma^2_j$ (with rounding
to the nearest integer).
These parameters can be estimated and tuned 
in various ways described in \cite{Koubi-Hill-Climbing}
and a followup work \cite{Koubi-Followup}.
We estimate them in an offline training procedure
where we find many DTSs with a larger scope $M'$ than the 
desired scope $M$ and a potentially
smaller number of rulers $n'$ than the desired number
$n$. This is done by running the 
algorithm with uniform mark sampling 
for these easier parameters and is fast to
do because bad DTSs are easy to find.
We then linearly scale the
sample means and standard deviations 
for the marks of the DTSs found
by a factor of $M/M'$ to get the
estimates needed for the scope of interest.

Lastly but importantly, 
we observe that there is a major limitation to
Algorithm \ref{fast-conditional-mark-insertion} when implemented in software which is the $64$-bit data widths of modern processors limiting us to $M + 1 = 64$. 
Larger data widths can be emulated by implementing
bitwise and shifting operations sequentially
on $64$-bit chunks with the shifting operations 
also requiring significant additional carryover operations. This is done in our
software implementation where we allow $M+1 \leq 256$.

\begin{algorithm}[t]
	\caption{Modified Koubi et al.\ \cite{Koubi-Hill-Climbing} algorithm.}\label{modified-Koubi}
	\begin{algorithmic}[1]
		\State initialize partial DTS rows with zero as the first mark
		\State $\mathsf{iters_1} \gets 0$
		\While{partial DTS incomplete and $\mathsf{iters_1} < \mathsf{thresh_1}$}
			\State $\mathsf{iters_1} \gets \mathsf{iters_1} + 1$
			\State $\mathsf{iters_2} \gets 0$
			\While{row incomplete and $\mathsf{iters_2} < \mathsf{thresh_2}$}
				\State $\mathsf{iters_2} \gets \mathsf{iters_2} + 1$
				\State attempt a mark insertion with Algorithm \ref{fast-conditional-mark-insertion} 
			\EndWhile
			\If{full row completed}
				\State insert row into partial DTS
			\Else
				\For{every completed row of partial DTS}
					\State remove completed row from partial DTS
					\State $\mathsf{iters_2} \gets 0$
					\While{row incomplete and $\mathsf{iters_2} < \mathsf{thresh_2}$}
					\State $\mathsf{iters_2} \gets \mathsf{iters_2} + 1$
					\State attempt a mark insertion with Algorithm \ref{fast-conditional-mark-insertion} 
					\EndWhile
					\If{full row completed}
						\State insert row into partial DTS
						\State \textbf{break}
					\Else
						\State undo removal of completed row
					\EndIf
				\EndFor
			\EndIf
		\EndWhile
	\end{algorithmic}
\end{algorithm}

\subsection{Hardware Implementation}\label{hardware-section}

\subsubsection{Mark Insertion}
In hardware, we can realize Algorithm \ref{fast-conditional-mark-insertion}
exactly without the limitations of datapath widths and with
all operations executed in a single clock cycle. However,
the variable shifting operations especially, as well as the non-bitwise
operations, require wide logic gates or large \emph{fan-in}. 
This
is realized by cascading smaller gates and results in large delays and thus a low maximum clock frequency. We instead pipeline this algorithm 
with the shifting being realized by pipelined trees 
of logarithmic depth in $M+1$ and other balanced delays 
inserted throughout to reduce fan-in. 
This results in a higher clock frequency
at the cost of an initial pipeline latency which is logarithmic in 
$M+1$.

Next, we note that the resulting pipeline has a feedback path
since the validity of an inserted mark depends on all previously inserted marks and thus the same state variables
that are being modified by the pipeline. This means that the pipeline must be cleared every time a mark is successfully inserted which incurs us
the aforementioned latency repeatedly. 
If every mark generated is valid, 
then the pipeline must be cleared as soon as
any mark traverses it which amounts to 
serialization and defeats the purpose of pipelining. 
However,
in the later stages of populating a DTS which is where most of the
algorithm time is spent, many mark insertions must be
attempted before finding a valid one and successfully modifying the DTS and thus changing the feedback path.
In this context, the pipeline flushes are infrequent enough to amortize the latencies and the overall speedup of the algorithm is almost equal to the relative increase in clock speed obtained by pipelining.
\subsubsection{Randomness Generation}
We obtain parameters for
our Gaussians offline with the software algorithm as described in Section \ref{software-section}. Since
these are rounded to integers, they are equivalent 
to discrete non-uniform distributions which
can be computed offline and generated
by accordingly binning a source of uniform randomness---also known as the 
inverse cumulative distribution function (CDF) method.

We get uniform pseudorandomness via maximum-length
linear-feedback shift registers (LFSRs) the theory of
which is developed by Golomb in \cite{Golomb-Shift-Register-Sequences}.
These are trivial
in hardware by construction. Moreover, we 
can implement the inverse CDF method 
with very low precision to obtain small circuits.
This has little impact on performance since 
the Gaussian model was an approximation of a
finite-support discrete distribution to begin
with and was only convenient in software due to
the dedicated floating-point units of modern
processors. We also note that in hardware, 
randomness generation 
is an independent, 
strictly feedforward process making it
trivial to pipeline and optimize 
independently of the rest of the system.

Lastly, when we wish to execute parallel
and independent searches for the same DTS 
on multiple FPGA devices 
with the same design, we can avoid the problem
of providing different seed values to each device by introducing true randomness via an entropy source. 
Many FPGA devices provide access
to their die temperature for thermal management 
via internal sensors. By occasionally XORing 
such temperature measurement data into the feedback
of the LFSRs, we make them non-deterministic.

\subsubsection{Complete System Design}
In addition to reducing fan-in, another principle of 
efficient hardware design is the reduction of \emph{fan-out} 
by distributing control as much as possible. Not
doing so results in routing congestion which 
prevents scaling the design to utilize
the full FPGA area.
This is done in our realization of Algorithm
\ref{modified-Koubi}. A simplified depiction of
the resulting architecture of a DTS search unit or 
\emph{worker} 
is provided in Figure ~\ref{hardware-figure}. The overall
design consists of a number of such parallel
workers together with a
system for selecting and outputting 
the result of the first worker 
who successfully finds a DTS.
Doing this efficiently and scalably 
also requires careful design.
The overall design was thoroughly
optimized for speed and scalability 
by applying the principles of reducing fan-in
and fan-out.
The complete details are available
in the implementation \cite{dts-search-hdl}.

The resulting system searches for an $(n,k)$-DTS of 
scope $M$ for specific fixed values of 
$n$, $k$, and $M$ that are 
not changeable at runtime since they directly
determine fundamental aspects of the underlying
hardware as well as the number of workers that
can be fit onto a given FPGA device. 
After eventually finding
such a DTS, the design signals completion, outputs
the DTS via a low speed serial transmission,
and becomes useless. The FPGA must then be
fully reconfigured to search for a
DTS with different parameters. This approach
thus fundamentally relies on the reprogrammability
of FPGAs.

\section{Future Work}\label{conclusion-section}
While we have provided a highly-effective approach to DTS
search with FPGAs, we have not attempted to exploit
the powerful algebra-aided approaches
surveyed in \cite{Shearer-IBM-Report}. Future work
should consider whether such approaches, and other
more intelligent heuristics in general, can be 
naturally fit into a digital design framework.
Secondly, we may consider whether other computationally
challenging problems arising in combinatorial design theory 
admit natural hardware approaches.

{\appendices
\section{Difference Triangle Sets}\label{DTS-appendix}

We exclude the first column of all DTSs that follow 
since it always consists of zeros.

\begin{gather*}
	\arraycolsep=2pt
	\begin{array}{cccc}
		2 & 80 & 119 & 140 \\
		30 & 53 & 111 & 139 \\
		27 & 61 & 125 & 137 \\
		4 & 46 & 101 & 136 \\
		20 & 69 & 95 & 135 \\
		8 & 51 & 87 & 134 \\
		31 & 68 & 116 & 133 \\
		24 & 38 & 120 & 131 \\
		9 & 63 & 108 & 130 \\
		25 & 41 & 114 & 129 \\
		5 & 57 & 127 & 128 \\
		19 & 32 & 91 & 124 \\
		18 & 62 & 112 & 118 \\
		29 & 103 & 106 & 113 \\
	\end{array}\;
	\arraycolsep=2pt
	\begin{array}{cccc}
		30 & 79 & 131 & 151 \\
		40 & 43 & 135 & 150 \\
		27 & 87 & 111 & 149 \\
		1 & 33 & 113 & 148 \\
		9 & 76 & 132 & 146 \\
		39 & 55 & 133 & 145 \\
		51 & 97 & 139 & 144 \\
		44 & 75 & 85 & 143 \\
		13 & 77 & 138 & 142 \\
		11 & 59 & 116 & 141 \\
		26 & 71 & 134 & 140 \\
		19 & 53 & 119 & 136 \\
		2 & 91 & 120 & 128 \\
		23 & 73 & 109 & 127 \\
		22 & 96 & 103 & 124 \\
	\end{array}\\[4pt]
	\arraycolsep=2pt
	\begin{array}{rrrrr}
		11 & 58 & 75 & 98 & 112 \\
		12 & 32 & 50 & 103 & 111 \\
		22 & 41 & 89 & 104 & 110 \\
		28 & 52 & 83 & 108 & 109 \\
		13 & 62 & 72 & 105 & 107 \\
		9 & 16 & 60 & 102 & 106 \\
		27 & 30 & 66 & 95 & 100 \\
	\end{array}\;
	\arraycolsep=2pt
	\begin{array}{rrrrr}
		14 & 49 & 85 & 122 & 128 \\
		11 & 23 & 77 & 105 & 127 \\
		3 & 45 & 100 & 109 & 126 \\
		30 & 63 & 87 & 121 & 125 \\
		13 & 44 & 83 & 103 & 124 \\
		1 & 48 & 53 & 113 & 120 \\
		16 & 18 & 92 & 102 & 117 \\
		19 & 46 & 75 & 107 & 115 \\
	\end{array}\\[4pt]
	\arraycolsep=2pt
	\begin{array}{rrrrr}
		33 & 55 & 76 & 142 & 145 \\
		11 & 47 & 115 & 128 & 144 \\
		5 & 42 & 72 & 137 & 143 \\
		28 & 63 & 103 & 121 & 141 \\
		24 & 26 & 80 & 126 & 140 \\
		19 & 34 & 98 & 108 & 139 \\
		44 & 51 & 83 & 135 & 136 \\
		48 & 57 & 107 & 130 & 134 \\
		12 & 61 & 106 & 123 & 131 \\
	\end{array}\;
	\arraycolsep=2pt
	\begin{array}{rrrrr}
		20 & 95 & 104 & 155 & 163 \\
		33 & 77 & 122 & 147 & 162 \\
		10 & 63 & 98 & 159 & 160 \\
		21 & 24 & 103 & 151 & 158 \\
		26 & 65 & 83 & 141 & 157 \\
		11 & 38 & 110 & 139 & 156 \\
		30 & 67 & 73 & 123 & 154 \\
		13 & 41 & 119 & 121 & 153 \\
		14 & 36 & 100 & 105 & 152 \\
		12 & 54 & 125 & 144 & 148 \\
	\end{array}\\[4pt]
	\arraycolsep=2pt
	\begin{array}{rrrrr}
		11 & 67 & 106 & 165 & 180 \\
		29 & 52 & 102 & 153 & 179 \\
		1 & 34 & 134 & 171 & 178 \\
		18 & 31 & 122 & 136 & 176 \\
		9 & 90 & 120 & 139 & 175 \\
		48 & 75 & 110 & 157 & 174 \\
		12 & 70 & 131 & 141 & 173 \\
		5 & 43 & 88 & 164 & 168 \\
		46 & 66 & 138 & 160 & 162 \\
		16 & 41 & 69 & 148 & 156 \\
		3 & 68 & 89 & 146 & 152 \\
	\end{array}\;
	\arraycolsep=2pt
	\begin{array}{rrrrr}
		10 & 83 & 125 & 179 & 196 \\
		27 & 91 & 136 & 189 & 195 \\
		39 & 63 & 99 & 160 & 193 \\
		26 & 44 & 106 & 155 & 192 \\
		20 & 52 & 120 & 171 & 190 \\
		43 & 47 & 122 & 157 & 188 \\
		5 & 7 & 139 & 147 & 187 \\
		9 & 93 & 116 & 174 & 185 \\
		3 & 25 & 127 & 156 & 184 \\
		34 & 55 & 137 & 167 & 183 \\
		1 & 77 & 90 & 164 & 178 \\
		12 & 50 & 117 & 158 & 173 \\
	\end{array}\\[4pt]
	\arraycolsep=2pt
	\begin{array}{rrrrr}
		12 & 95 & 104 & 206 & 213 \\
		8 & 114 & 128 & 172 & 212 \\
		29 & 89 & 144 & 168 & 211 \\
		30 & 48 & 147 & 188 & 210 \\
		25 & 64 & 133 & 190 & 209 \\
		17 & 51 & 138 & 170 & 208 \\
		20 & 56 & 130 & 205 & 207 \\
		46 & 73 & 134 & 169 & 202 \\
		1 & 72 & 137 & 196 & 199 \\
		6 & 91 & 177 & 181 & 192 \\
		16 & 37 & 82 & 179 & 189 \\
		28 & 78 & 131 & 178 & 183 \\
		26 & 49 & 80 & 161 & 174 \\
	\end{array}\;
	\arraycolsep=2pt
	\begin{array}{rrrrr}
		42 & 70 & 124 & 214 & 230 \\
		21 & 89 & 148 & 189 & 229 \\
		27 & 62 & 142 & 220 & 228 \\
		4 & 53 & 101 & 167 & 227 \\
		23 & 88 & 135 & 206 & 226 \\
		13 & 43 & 116 & 218 & 225 \\
		32 & 99 & 111 & 209 & 224 \\
		25 & 130 & 159 & 161 & 222 \\
		26 & 72 & 146 & 157 & 221 \\
		57 & 76 & 180 & 213 & 219 \\
		1 & 51 & 95 & 179 & 216 \\
		38 & 52 & 185 & 202 & 207 \\
		18 & 87 & 170 & 194 & 204 \\
		3 & 58 & 154 & 190 & 199 \\
	\end{array}\\[4pt]
	\arraycolsep=2pt
	\begin{array}{rrrrr}
		54 & 100 & 136 & 234 & 249 \\
		9 & 37 & 133 & 206 & 248 \\
		11 & 87 & 152 & 230 & 247 \\
		56 & 57 & 165 & 213 & 246 \\
		62 & 69 & 159 & 243 & 245 \\
		23 & 66 & 177 & 228 & 244 \\
		26 & 38 & 130 & 224 & 242 \\
		3 & 77 & 170 & 202 & 241 \\
		20 & 47 & 119 & 235 & 240 \\
		19 & 102 & 203 & 233 & 237 \\
		24 & 103 & 147 & 192 & 232 \\
		52 & 60 & 200 & 210 & 231 \\
		25 & 75 & 163 & 212 & 226 \\
		29 & 120 & 155 & 161 & 225 \\
		22 & 80 & 139 & 194 & 207 \\
	\end{array}\;
	\arraycolsep=2pt
	\begin{array}{rrrrrr}
		22 & 48 & 55 & 99 & 105 & 118 \\
		10 & 45 & 75 & 92 & 103 & 117 \\
		12 & 16 & 43 & 80 & 114 & 116 \\
		5 & 20 & 66 & 74 & 106 & 115 \\
		1 & 24 & 53 & 91 & 109 & 112 \\
	\end{array}\\[4pt]
	\arraycolsep=2pt
	\begin{array}{rrrrrr}
		12 & 42 & 104 & 107 & 142 & 144 \\
		5 & 19 & 64 & 72 & 127 & 143 \\
		23 & 27 & 97 & 121 & 128 & 141 \\
		15 & 58 & 84 & 106 & 131 & 140 \\
		17 & 49 & 85 & 126 & 136 & 137 \\
		18 & 46 & 75 & 96 & 129 & 135 \\
	\end{array}\;
	\arraycolsep=2pt
	\begin{array}{rrrrrr}
		2 & 33 & 91 & 92 & 144 & 170 \\
		16 & 40 & 85 & 97 & 140 & 169 \\
		6 & 41 & 73 & 147 & 155 & 166 \\
		20 & 50 & 88 & 116 & 152 & 165 \\
		4 & 27 & 98 & 107 & 154 & 161 \\
		21 & 46 & 108 & 122 & 156 & 159 \\
		22 & 39 & 83 & 143 & 148 & 158 \\
	\end{array}\\[4pt]
	\arraycolsep=2pt
	\begin{array}{rrrrrr}
		2 & 69 & 118 & 123 & 182 & 196 \\
		24 & 61 & 108 & 133 & 183 & 193 \\
		1 & 30 & 87 & 129 & 175 & 192 \\
		4 & 26 & 96 & 161 & 170 & 189 \\
		16 & 34 & 110 & 131 & 176 & 187 \\
		12 & 32 & 83 & 136 & 151 & 184 \\
		3 & 38 & 44 & 150 & 158 & 181 \\
		39 & 52 & 79 & 134 & 141 & 177 \\
	\end{array}\;
	\arraycolsep=2pt
	\begin{array}{rrrrrr}
		20 & 110 & 128 & 172 & 215 & 221 \\
		11 & 66 & 89 & 137 & 191 & 219 \\
		7 & 42 & 134 & 149 & 217 & 218 \\
		9 & 76 & 155 & 182 & 212 & 216 \\
		12 & 52 & 85 & 183 & 200 & 214 \\
		16 & 74 & 113 & 177 & 190 & 209 \\
		38 & 41 & 94 & 159 & 185 & 206 \\
		8 & 37 & 88 & 151 & 187 & 197 \\
		25 & 70 & 75 & 170 & 192 & 194 \\
	\end{array}\\[4pt]
	\arraycolsep=2pt
	\begin{array}{rrrrrr}
		59 & 71 & 141 & 216 & 245 & 247 \\
		16 & 61 & 111 & 158 & 209 & 244 \\
		17 & 58 & 151 & 177 & 207 & 243 \\
		6 & 87 & 115 & 152 & 200 & 242 \\
		22 & 94 & 161 & 195 & 214 & 241 \\
		38 & 114 & 117 & 222 & 235 & 240 \\
		14 & 63 & 102 & 154 & 227 & 238 \\
		4 & 100 & 107 & 169 & 229 & 237 \\
		20 & 43 & 64 & 199 & 223 & 232 \\
		1 & 78 & 132 & 206 & 221 & 231 \\
	\end{array}\;
	\arraycolsep=2pt
	\begin{array}{rrrrrr}
		13 & 98 & 127 & 224 & 227 & 273 \\
		16 & 104 & 123 & 200 & 234 & 272 \\
		21 & 43 & 124 & 230 & 265 & 271 \\
		57 & 105 & 122 & 223 & 259 & 270 \\
		7 & 37 & 99 & 179 & 242 & 269 \\
		52 & 108 & 113 & 177 & 248 & 268 \\
		1 & 75 & 194 & 208 & 239 & 267 \\
		15 & 54 & 182 & 186 & 252 & 264 \\
		2 & 60 & 153 & 176 & 255 & 263 \\
		25 & 76 & 109 & 159 & 245 & 254 \\
		10 & 42 & 131 & 199 & 225 & 243 \\
	\end{array}\\[4pt]
	\arraycolsep=2pt
	\begin{array}{rrrrrr}
		44 & 107 & 190 & 204 & 281 & 298 \\
		22 & 42 & 126 & 229 & 288 & 297 \\
		57 & 96 & 163 & 176 & 211 & 296 \\
		36 & 122 & 141 & 220 & 271 & 294 \\
		30 & 32 & 169 & 218 & 247 & 293 \\
		50 & 76 & 197 & 253 & 278 & 290 \\
		27 & 60 & 161 & 216 & 270 & 286 \\
		15 & 62 & 102 & 194 & 260 & 284 \\
		7 & 71 & 143 & 238 & 272 & 283 \\
		41 & 99 & 164 & 264 & 274 & 282 \\
		1 & 53 & 114 & 249 & 252 & 280 \\
		6 & 88 & 157 & 230 & 268 & 273 \\
	\end{array}\;
	\arraycolsep=2pt
	\begin{array}{rrrrrr}
		16 & 46 & 147 & 215 & 268 & 322 \\
		35 & 63 & 143 & 242 & 302 & 321 \\
		31 & 92 & 157 & 259 & 279 & 318 \\
		24 & 47 & 162 & 244 & 313 & 317 \\
		22 & 110 & 204 & 240 & 249 & 316 \\
		32 & 33 & 173 & 245 & 297 & 315 \\
		12 & 50 & 116 & 250 & 307 & 312 \\
		6 & 95 & 103 & 247 & 284 & 311 \\
		2 & 93 & 174 & 225 & 303 & 310 \\
		56 & 133 & 146 & 251 & 299 & 309 \\
		43 & 83 & 127 & 233 & 275 & 304 \\
		25 & 74 & 170 & 185 & 211 & 298 \\
		11 & 86 & 120 & 274 & 288 & 291 \\
	\end{array}\\[4pt]
	\arraycolsep=2pt
	\begin{array}{rrrrrr}
		6 & 84 & 172 & 315 & 347 & 351 \\
		24 & 96 & 205 & 294 & 337 & 350 \\
		60 & 113 & 124 & 295 & 316 & 349 \\
		41 & 50 & 144 & 261 & 296 & 348 \\
		61 & 119 & 142 & 250 & 327 & 343 \\
		31 & 121 & 195 & 237 & 334 & 336 \\
		49 & 63 & 210 & 249 & 320 & 335 \\
		19 & 137 & 155 & 228 & 311 & 333 \\
		44 & 51 & 227 & 265 & 302 & 331 \\
		20 & 82 & 150 & 197 & 304 & 330 \\
		55 & 85 & 112 & 223 & 324 & 329 \\
		3 & 98 & 165 & 288 & 300 & 328 \\
		1 & 46 & 80 & 194 & 253 & 323 \\
		10 & 102 & 167 & 293 & 301 & 318 \\
	\end{array}\;
	\arraycolsep=2pt
	\begin{array}{rrrrrr}
		73 & 109 & 161 & 274 & 358 & 373 \\
		11 & 68 & 174 & 290 & 333 & 372 \\
		70 & 153 & 156 & 313 & 346 & 371 \\
		35 & 61 & 223 & 255 & 347 & 370 \\
		48 & 150 & 178 & 332 & 345 & 369 \\
		40 & 90 & 248 & 294 & 360 & 368 \\
		17 & 128 & 224 & 231 & 357 & 366 \\
		56 & 125 & 187 & 192 & 324 & 362 \\
		30 & 101 & 241 & 272 & 282 & 359 \\
		53 & 54 & 129 & 239 & 334 & 356 \\
		6 & 55 & 206 & 257 & 350 & 354 \\
		64 & 80 & 232 & 253 & 351 & 353 \\
		47 & 89 & 196 & 216 & 310 & 339 \\
		14 & 155 & 240 & 259 & 319 & 331 \\
		63 & 81 & 108 & 247 & 291 & 325 \\
	\end{array}\\[4pt]
	\arraycolsep=2pt
	\begin{array}{rrrrrrr}
		5 & 43 & 49 & 82 & 119 & 123 & 134 \\
		17 & 25 & 65 & 96 & 120 & 122 & 132 \\
		20 & 23 & 50 & 84 & 112 & 113 & 131 \\
		9 & 22 & 54 & 68 & 75 & 110 & 126 \\
	\end{array}
	\;
	\arraycolsep=2pt
	\begin{array}{rrrrrrr}
		1 & 19 & 58 & 71 & 139 & 143 & 170 \\
		2 & 24 & 47 & 91 & 103 & 131 & 168 \\
		6 & 21 & 75 & 108 & 116 & 158 & 167 \\
		17 & 53 & 60 & 115 & 150 & 153 & 164 \\
		5 & 34 & 66 & 114 & 130 & 140 & 160 \\
	\end{array}
	\\[4pt]
	\arraycolsep=2pt
	\begin{array}{rrrrrrr}
		12 & 41 & 47 & 125 & 146 & 204 & 208 \\
		14 & 60 & 104 & 142 & 179 & 205 & 207 \\
		40 & 51 & 71 & 137 & 173 & 189 & 206 \\
		1 & 43 & 73 & 96 & 153 & 187 & 202 \\
		10 & 18 & 74 & 141 & 150 & 195 & 200 \\
		22 & 70 & 77 & 109 & 170 & 194 & 197 \\
	\end{array}
	\\[4pt]
	\arraycolsep=2pt
	\begin{array}{rrrrrrr}
		10 & 28 & 69 & 184 & 199 & 235 & 249 \\
		1 & 53 & 113 & 161 & 203 & 223 & 248 \\
		31 & 68 & 138 & 154 & 209 & 236 & 243 \\
		21 & 64 & 122 & 139 & 185 & 215 & 241 \\
		2 & 40 & 49 & 144 & 183 & 216 & 240 \\
		4 & 120 & 128 & 131 & 208 & 231 & 237 \\
		13 & 32 & 67 & 146 & 158 & 219 & 224 \\
	\end{array}
	\\[4pt]
	\arraycolsep=2pt
	\begin{array}{rrrrrrr}
		23 & 59 & 115 & 175 & 224 & 237 & 281 \\
		7 & 46 & 117 & 143 & 216 & 261 & 280 \\
		28 & 61 & 75 & 159 & 189 & 263 & 279 \\
		1 & 69 & 135 & 157 & 265 & 268 & 276 \\
		2 & 34 & 87 & 127 & 232 & 259 & 274 \\
		35 & 89 & 113 & 190 & 262 & 266 & 271 \\
		50 & 67 & 79 & 174 & 229 & 260 & 270 \\
		21 & 58 & 123 & 206 & 226 & 244 & 269 \\
	\end{array}
	\\[4pt]
	\arraycolsep=2pt
	\begin{array}{rrrrrrr}
		8 & 82 & 92 & 203 & 281 & 315 & 317 \\
		46 & 89 & 102 & 218 & 258 & 313 & 316 \\
		16 & 47 & 88 & 157 & 222 & 288 & 308 \\
		44 & 63 & 108 & 125 & 257 & 299 & 306 \\
		1 & 61 & 167 & 179 & 246 & 276 & 305 \\
		15 & 24 & 148 & 183 & 284 & 298 & 302 \\
		38 & 71 & 158 & 226 & 248 & 280 & 301 \\
		27 & 100 & 123 & 180 & 217 & 265 & 293 \\
		26 & 51 & 103 & 186 & 279 & 285 & 290 \\
	\end{array}
	\\[4pt]
	\arraycolsep=2pt
	\begin{array}{rrrrrrr}
		34 & 95 & 145 & 219 & 238 & 348 & 351 \\
		45 & 65 & 77 & 242 & 272 & 345 & 350 \\
		24 & 94 & 149 & 178 & 287 & 340 & 349 \\
		39 & 79 & 96 & 240 & 265 & 308 & 346 \\
		66 & 82 & 133 & 222 & 274 & 297 & 343 \\
		31 & 49 & 184 & 283 & 320 & 330 & 341 \\
		35 & 48 & 76 & 218 & 224 & 336 & 338 \\
		33 & 92 & 172 & 180 & 244 & 331 & 335 \\
		54 & 98 & 214 & 228 & 311 & 318 & 333 \\
		1 & 86 & 128 & 188 & 259 & 295 & 322 \\
	\end{array}
	\\[4pt]
	\arraycolsep=2pt
	\begin{array}{rrrrrrr}
		7 & 61 & 117 & 191 & 271 & 376 & 391 \\
		43 & 100 & 196 & 262 & 295 & 335 & 390 \\
		50 & 53 & 138 & 326 & 352 & 371 & 389 \\
		1 & 77 & 198 & 240 & 242 & 375 & 388 \\
		32 & 46 & 136 & 175 & 268 & 349 & 380 \\
		28 & 38 & 178 & 199 & 344 & 350 & 379 \\
		64 & 111 & 115 & 234 & 342 & 358 & 378 \\
		60 & 68 & 186 & 284 & 343 & 368 & 373 \\
		58 & 141 & 152 & 164 & 266 & 345 & 367 \\
		9 & 87 & 158 & 255 & 289 & 337 & 364 \\
		24 & 65 & 137 & 207 & 293 & 310 & 362 \\
	\end{array}
	\\[4pt]
	\arraycolsep=2pt
	\begin{array}{rrrrrrr}
		8 & 65 & 252 & 289 & 408 & 421 & 430 \\
		29 & 72 & 177 & 239 & 328 & 419 & 425 \\
		31 & 58 & 154 & 271 & 313 & 407 & 424 \\
		47 & 81 & 160 & 231 & 368 & 401 & 420 \\
		39 & 88 & 284 & 302 & 316 & 406 & 418 \\
		1 & 46 & 139 & 195 & 305 & 358 & 417 \\
		16 & 115 & 130 & 288 & 380 & 391 & 415 \\
		23 & 101 & 229 & 232 & 350 & 404 & 411 \\
		63 & 83 & 147 & 207 & 280 & 405 & 409 \\
		21 & 106 & 108 & 183 & 329 & 398 & 403 \\
		10 & 51 & 191 & 267 & 334 & 362 & 402 \\
		50 & 157 & 205 & 243 & 325 & 369 & 395 \\
	\end{array}
	\\[4pt]
	\arraycolsep=2pt
	\begin{array}{rrrrrrr}
		11 & 60 & 167 & 254 & 276 & 404 & 461 \\
		42 & 163 & 221 & 320 & 373 & 440 & 459 \\
		80 & 176 & 226 & 289 & 368 & 443 & 458 \\
		101 & 131 & 166 & 271 & 406 & 449 & 457 \\
		1 & 93 & 223 & 248 & 436 & 438 & 454 \\
		55 & 83 & 199 & 335 & 382 & 419 & 452 \\
		46 & 118 & 141 & 292 & 380 & 441 & 451 \\
		39 & 162 & 203 & 251 & 428 & 431 & 448 \\
		76 & 82 & 184 & 332 & 388 & 442 & 447 \\
		31 & 52 & 129 & 143 & 301 & 433 & 446 \\
		24 & 138 & 211 & 285 & 379 & 411 & 445 \\
		26 & 64 & 104 & 201 & 328 & 372 & 434 \\
		29 & 36 & 218 & 227 & 351 & 420 & 432 \\
	\end{array}
	\\[4pt]
	\arraycolsep=2pt
	\begin{array}{rrrrrrr}
		58 & 83 & 229 & 420 & 479 & 487 & 498 \\
		12 & 120 & 233 & 289 & 443 & 456 & 494 \\
		32 & 87 & 124 & 227 & 371 & 473 & 493 \\
		64 & 145 & 179 & 286 & 462 & 464 & 492 \\
		43 & 100 & 237 & 354 & 424 & 485 & 489 \\
		80 & 94 & 257 & 361 & 382 & 459 & 488 \\
		27 & 153 & 243 & 393 & 475 & 481 & 484 \\
		31 & 147 & 201 & 329 & 418 & 465 & 480 \\
		53 & 105 & 215 & 260 & 412 & 460 & 478 \\
		22 & 121 & 181 & 304 & 436 & 472 & 477 \\
		24 & 138 & 184 & 280 & 364 & 437 & 476 \\
		35 & 75 & 209 & 310 & 395 & 458 & 474 \\
		44 & 86 & 219 & 314 & 453 & 463 & 470 \\
		76 & 109 & 274 & 348 & 417 & 466 & 467 \\
	\end{array}
	\\[4pt]
	\arraycolsep=2pt
	\begin{array}{rrrrrrr}
		20 & 38 & 177 & 338 & 502 & 516 & 523 \\
		45 & 108 & 135 & 314 & 455 & 510 & 521 \\
		54 & 128 & 171 & 316 & 436 & 488 & 520 \\
		28 & 150 & 248 & 252 & 451 & 490 & 519 \\
		61 & 114 & 154 & 344 & 459 & 475 & 517 \\
		57 & 103 & 183 & 255 & 415 & 509 & 514 \\
		36 & 67 & 142 & 218 & 365 & 444 & 513 \\
		41 & 64 & 153 & 285 & 394 & 396 & 512 \\
		10 & 175 & 212 & 263 & 313 & 443 & 508 \\
		1 & 163 & 236 & 292 & 429 & 500 & 506 \\
		34 & 119 & 215 & 328 & 487 & 495 & 504 \\
		22 & 127 & 213 & 273 & 442 & 489 & 501 \\
		82 & 95 & 205 & 379 & 405 & 483 & 486 \\
		35 & 97 & 222 & 246 & 432 & 447 & 480 \\
		25 & 44 & 168 & 304 & 334 & 425 & 474 \\
	\end{array}
\end{gather*}

\section{Device Details and Performance Estimates}\label{engineering-appendix}
The searches were primarily performed on a collection of six to eight independent Xilinx Kintex 7 XC7K325T FPGAs, which are mid-range 28 nm FPGAs from around 2012. The designs ran at 
250 to 300 MHz at about 75\% to 90\% utilization
of the device with power draws of around 
15 to 18 W per device.
The majority of cases were run at 300 MHz with 80\%
utilization which are very high numbers 
for this class of device.
The average time taken to find each DTS
was about 4 to 5 days. 

Some searches
were also done
with an Altera Stratix IV EP4SE820 FPGA 
which is a high-end 40 nm FPGA from around 2009.
This device was found to fit about twice the
number of workers as the Kintex 7 but at half 
of the clock frequency making it about equivalent
for most cases. A notable example is the search for
the $(5,7)$-DTS of scope $170$ where a design with
77 workers achieved 150 MHz at 99\% utilization of
the Stratix IV. 
The core design makes no use
of specialized hardened FPGA resources other than inferred generic
embedded RAMs making it easy to port to different devices.
This combined with the ability to scale to fully utilize
any FPGA while executing a useful and complex task
makes it an interesting benchmark of generic 
FPGA capabilities.

Quantifying performance gains
for this style of problem is somewhat tricky
since we are not so much concerned with how many 
$(n,k)$-DTSs of scope $M$ for fixed values of 
$n$, $k$, and $M$, can be found in a given 
amount of time, but with whether a single such DTS
can be found at all in any reasonable amount of time.
Moreover, the speed with which sub-optimal DTSs 
can be found repeatedly is not reflective of 
the speed with which a 
single optimal or near-optimal DTS can be found. 
This is because a method can be tuned to
find sub-optimal DTSs quickly 
by imposing significant structure
on the search space which could impede 
the same method from finding an optimal DTS at all.
With these fine points in mind, 
we provide some numerical figures.

A software-based search for the optimal $(14,4)$-DTS
(with scope $140$) was run on a single
AMD Ryzen 9 7950X CPU 
which is a 16-core, 5 nm processor from around 2022
which runs at around 5 GHz. 
The search used $12$ parallel instances and the 
CPU was otherwise 
occasionally 
loaded with unrelated single-threaded tasks.
The search
was unsuccessful after over 70 days and
the estimated attributed power draw was
about 130 W throughout.
The same software algorithm was successful at
finding the optimal 
$(12,4)$-DTS and $(13,4)$-DTS
in \cite{hosc} in less time per DTS.
An FPGA-based search was 
then run on a \emph{single} Kintex 7
FPGA fitting 48 workers 
and found the optimal $(14,4)$-DTS in 10 days.
The design in this case ran
at 300 MHz with 80\% utilization
drawing around 16 W. 

Next, we attempt a more reliable quantification 
of the CPU versus FPGA performance 
with the aforementioned caveats in mind. 
We consider the problem of finding 
a $(16,4)$-DTSs of scope $165$ (which is sub-optimal)
repeatedly and measure the time
taken to find such a DTS. 
We run $32$ instances
of the software-based search on
a single AMD Ryzen 9 7950X3D CPU which is
distinct from but very similar to the 7950X. 
We then also search for DTSs with the same parameters
using a maximally utilized single Kintex 7 FPGA.
In both cases, we measure the time taken until
the first success. This was repeated 10 times 
and the average time taken by the CPU
was around 1100 seconds with the average time
taken by the FPGA being around 150 seconds.
The estimated power draws are around
120 W for the CPU and 16 W for the FPGA.
From a cost perspective, salvaged Kintex 7
devices (which we used) are available at around one-fifth 
of the cost of the 5 nm, 16-core CPUs considered, excluding the
much more expensive peripherals 
(motherboards, memories, power supplies, and cooling solutions)
needed to use such CPUs. This yields
a rough estimate of seven times the performance
at one-seventh of the power and one-fifth of the cost.

We conclude with an 
important caveat: While our software baseline
was highly-optimized, this does not at all 
preclude the possibility that 
a better implementation or a 
different software-based algorithm altogether
could outperform our FPGA-based approach. 
The implementation details, code, 
and computing resources used for many past 
DTS searches are not published,
making it difficult to definitively establish
the relative effectiveness of our method.
However, our approach ultimately does 
beat existing DTS records and does 
so with old FPGAs, low cost, and low energy.
}


\begin{thebibliography}{10}
	\providecommand{\url}[1]{#1}
	\csname url@samestyle\endcsname
	\providecommand{\newblock}{\relax}
	\providecommand{\bibinfo}[2]{#2}
	\providecommand{\BIBentrySTDinterwordspacing}{\spaceskip=0pt\relax}
	\providecommand{\BIBentryALTinterwordstretchfactor}{4}
	\providecommand{\BIBentryALTinterwordspacing}{\spaceskip=\fontdimen2\font plus
		\BIBentryALTinterwordstretchfactor\fontdimen3\font minus
		\fontdimen4\font\relax}
	\providecommand{\BIBforeignlanguage}[2]{{%
			\expandafter\ifx\csname l@#1\endcsname\relax
			\typeout{** WARNING: IEEEtran.bst: No hyphenation pattern has been}%
			\typeout{** loaded for the language `#1'. Using the pattern for}%
			\typeout{** the default language instead.}%
			\else
			\language=\csname l@#1\endcsname
			\fi
			#2}}
	\providecommand{\BIBdecl}{\relax}
	\BIBdecl
	
	\bibitem{Klove}
	T.~Kl{\o}ve, ``Bounds and construction for difference triangle sets,''
	\emph{{IEEE} Trans. Inf. Theory}, vol.~35, no.~4, pp. 879--886, Jul. 1989.
	
	\bibitem{Klove-Bounds}
	------, ``Bounds on the size of optimal difference triangle sets,''
	\emph{{IEEE} Trans. Inf. Theory}, vol.~34, no.~2, pp. 355--361, Mar. 1988.
	
	\bibitem{Colbourn-IT}
	Y.~M. Chee and C.~J. Colbourn, ``Constructions for difference triangle sets,''
	\emph{{IEEE} Trans. Inf. Theory}, vol.~43, no.~4, pp. 1346--1349, Jul. 1997.
	
	\bibitem{Zhi-DDS}
	Z.~Chen, P.~Fan, and F.~Jin, ``Disjoint difference sets, difference triangle
	sets, and related codes,'' \emph{{IEEE} Trans. Inf. Theory}, vol.~38, no.~2,
	pp. 518--522, Mar. 1992.
	
	\bibitem{Chen-Further}
	Z.~Chen, ``Further results on difference triangle sets,'' \emph{{IEEE} Trans.
		Inf. Theory}, vol.~40, no.~4, pp. 1268--1270, Jul. 1994.
	
	\bibitem{DTS-Affine}
	A.~C.~H. Ling, ``Difference triangle sets from affine planes,'' \emph{{IEEE}
		Trans. Inf. Theory}, vol.~48, no.~8, pp. 2399--2401, Aug. 2002.
	
	\bibitem{Koubi-Hill-Climbing}
	S.~Koubi, M.~Mata-Montero, and N.~Shalaby, ``Using directed hill-climbing for
	the construction of difference triangle sets,'' \emph{{IEEE} Trans. Inf.
		Theory}, vol.~51, no.~1, pp. 335--339, Jan. 2005.
	
	\bibitem{MILP-DTS}
	R.~Lorentzen and R.~Nilsen, ``Application of linear programming to the optimal
	difference triangle set problem,'' \emph{{IEEE} Trans. Inf. Theory}, vol.~37,
	no.~5, pp. 1486--1488, Sep. 1991.
	
	\bibitem{Some-DTS}
	E.~J. Klieber, ``Some difference triangles for constructing self-orthogonal
	codes (corresp.),'' \emph{{IEEE} Trans. Inf. Theory}, vol.~16, no.~2, pp.
	237--238, Mar. 1970.
	
	\bibitem{CSOC}
	J.~Robinson and A.~Bernstein, ``A class of binary recurrent codes with limited
	error propagation,'' \emph{{IEEE} Trans. Inf. Theory}, vol.~13, no.~1, pp.
	106--113, Jan. 1967.
	
	\bibitem{NB-DTS}
	G.~N. Alfarano, J.~Lieb, and J.~Rosenthal, ``Construction of {LDPC}
	convolutional codes via difference triangle sets,'' \emph{Des., Codes
		Cryptography"}, vol.~89, pp. 2235--2254, 2021.
	
	\bibitem{CC-DTS}
	Y.~M. Chee, S.~H. Dau, A.~C.~H. Ling, and S.~Ling, ``Linear size optimal
	$q$-ary constant-weight codes and constant-composition codes,'' \emph{{IEEE}
		Trans. Inf. Theory}, vol.~56, no.~1, pp. 140--151, Jan. 2010.
	
	\bibitem{hosc}
	M.~Shehadeh, F.~R. Kschischang, A.~Y. Sukmadji, and W.~Kingsford,
	``Higher-order staircase codes,'' \emph{{IEEE} Trans. Inf. Theory}, early
	access, Feb.~20, 2025, doi: 10.1109/TIT.2025.3544168.
	
	\bibitem{OOC}
	F.~R.~K. Chung, J.~A. Salehi, and V.~K. Wei, ``Optical orthogonal codes:
	Design, analysis, and applications,'' \emph{{IEEE} Trans. Inf. Theory},
	vol.~35, no.~3, pp. 595--604, May 1989.
	
	\bibitem{DTS-SOOC}
	W.~Chu and S.~W. Golomb, ``A note on the equivalence between strict optical
	orthogonal codes and difference triangle sets,'' \emph{{IEEE} Trans. Inf.
		Theory}, vol.~49, no.~3, pp. 759--761, Mar. 2003.
	
	\bibitem{SOOC}
	J.-G. Zhang, ``Design of a special family of optical {CDMA} address codes for
	fully asynchronous data communications,'' \emph{{IEEE} Trans. Commun.},
	vol.~47, no.~7, pp. 967--973, Jul. 1999.
	
	\bibitem{Bermond}
	J.-C. Bermond, ``Graceful graphs, radio antennae and {French} windmills,'' in
	\emph{Proc. One Day Combinatorics Conf.}, ser. Res. Notes Math.,
	vol.~34.\hskip 1em plus 0.5em minus 0.4em\relax Pitman, 1979, pp. 18--37.
	
	\bibitem{Golomb-Ruler}
	G.~S. Bloom and S.~W. Golomb, ``Applications of numbered undirected graphs,''
	\emph{Proc.~IEEE}, vol.~65, no.~4, pp. 562--570, Apr. 1977.
	
	\bibitem{Babcock}
	W.~C. Babcock, ``Intermodulation interference in radio systems frequency of
	occurrence and control by channel selection,'' \emph{Bell Syst.~Tech.~J.},
	vol.~32, no.~1, pp. 63--73, Jan. 1953.
	
	\bibitem{distributed-OGR}
	``{distributed.net}: Project {OGR},'' \url{https://www.distributed.net/OGR},
	Nov. 2022.
	
	\bibitem{Apostol-Golomb}
	A.~Dimitromanolakis, ``Analysis of the {Golomb} ruler and the {Sidon} set
	problems, and determination of large, near-optimal {Golomb} rulers,'' Diploma
	Thesis, Technical University of Crete, 2002.
	
	\bibitem{Skolem}
	T.~Skolem, ``On certain distributions of integers in pairs with given
	differences,'' \emph{Mathematica Scandinavica}, vol.~5, pp. 57--68, 1957.
	
	\bibitem{OKeefe}
	E.~S. O'Keefe, ``Verification of a conjecture of {Th.~Skolem},''
	\emph{Mathematica Scandinavica}, vol.~9, pp. 80--82, 1961.
	
	\bibitem{Wild}
	P.~Wild, ``Combining perfect systems of difference sets,'' \emph{Bull. London
		Math. Soc.}, vol.~18, no.~2, pp. 127--131, 1986.
	
	\bibitem{Kotzig-Turgeon}
	A.~Kotzig and J.~M. Turgeon, ``Perfect systems of difference sets and additive
	sequences of permutations,'' in \emph{Proc. 10th South-Eastern Conf.
		Combinatorics, Graph Theory Comput.}, Boca Raton, FL, USA, Apr. 1979, pp.
	629--636.
	
	\bibitem{deg-4-1000}
	G.~Ge, Y.~Miao, and X.~Sun, ``Perfect difference families, perfect difference
	matrices, and related combinatorial structures,'' \emph{J. Combinatorial
		Des.}, vol.~18, no.~6, pp. 415--449, Nov. 2010.
	
	\bibitem{Shearer-LP}
	J.~B. Shearer, ``Improved {LP} lower bounds for difference triangle sets,''
	\emph{Electron.~J.~Combinatorics}, vol.~6, no. R31, 1999.
	
	\bibitem{Shearer-IBM-Programs-Paper}
	------, ``Some new difference triangle sets,'' \emph{J. Combinatorial Math.
		Combinatorial Comput.}, vol.~27, pp. 65--76, 1998.
	
	\bibitem{Shearer-IBM-Programs}
	------, ``Difference triangle set programs,'' Archived copy:
	\url{https://web.archive.org/web/20050318171624/http://www.research.ibm.com/people/s/shearer/dtsprog.html},
	2005 snapshot.
	
	\bibitem{TABU}
	P.~Galinier and B.~Jaumard, ``A tabu search algorithm for difference triangle
	sets and {Golomb} rulers,'' \emph{Comput.~Operations Res.}, vol.~33, no.~4,
	pp. 955--970, Apr. 2006.
	
	\bibitem{Koubi-Followup}
	S.~Koubi and N.~Shalaby, ``The combined use of a genetic algorithm and the
	hill-climbing algorithm to find difference triangle sets,'' \emph{J.
		Combinatorial Math. Combinatorial Comput.}, vol.~66, pp. 289--296, 2008.
	
	\bibitem{Shearer-IBM-Report}
	J.~B. Shearer, ``Difference triangle set constructions,'' IBM Res. Div., Tech.
	Rep. RC24623, Aug. 2008.
	
	\bibitem{Colbourn}
	C.~J. Colbourn and J.~H. Dinitz, Eds., \emph{Handbook of Combinatorial
		Designs}.\hskip 1em plus 0.5em minus 0.4em\relax Chapman \& Hall/CRC, 2007.
	
	\bibitem{dts-search-hdl}
	M.~Shehadeh, ``dts-search-hdl,'' GitHub repository:
	\url{https://github.com/applecoffeecake/dts-search-hdl}, Feb. 2025.
	
	\bibitem{Brown-Vranesic}
	S.~Brown and Z.~Vranesic, \emph{Fundamentals of Digital Logic wih Verilog
		Design}, 3rd~ed.\hskip 1em plus 0.5em minus 0.4em\relax {McGraw-Hill}, 2014.
	
	\bibitem{Kuon-Rose}
	I.~Kuon and J.~Rose, ``Measuring the gap between {FPGA}s and {ASIC}s,''
	\emph{{IEEE} Trans. Comput.-Aided Design Integr. Circuits Syst.}, vol.~26,
	no.~2, pp. 203--215, 2007.
	
	\bibitem{C-language}
	{ISO/IEC JTC1/SC22/WG14}, ``{ISO/IEC 9899:201x}: Programming languages ---
	{C},'' \url{https://www.open-std.org/jtc1/sc22/wg14/www/docs/n1570.pdf}, Apr.
	2011.
	
	\bibitem{Verilog}
	``{IEEE} standard for {SystemVerilog}--unified hardware design, specification,
	and verification language,'' \emph{IEEE Std 1800-2023 (Revision of IEEE Std
		1800-2017)}, 2024.
	
	\bibitem{Golomb-Shift-Register-Sequences}
	S.~W. Golomb, \emph{Shift Register Sequences}.\hskip 1em plus 0.5em minus
	0.4em\relax Holden-Day, 1967.
	
\end{thebibliography}


\end{document}